\documentclass[twocolumn,showpacs,aps,prl,superscriptaddress,floatfix]{revtex4}
\usepackage{graphicx}
\usepackage{verbatim}
\usepackage{dcolumn}
\usepackage{amsmath}
\usepackage{epsfig}

\newcommand{\BaBarYear}      {18}
\newcommand{\BaBarNumber}    {009}
\newcommand{\BaBarType}      {PUB}  
\newcommand{\SLACPubNumber}  {17335}

\input babarsym.tex

\def\epem{e^+e^-}
\def\MeV{\mev}
\def\GeV{\gev}

\def\figurebox#1#2#3{%
    \def\arg{#3}%
    \ifx\arg\empty
    {\hfill\vbox{\hsize#2\hrule\hbox to #2{\vrule\hfill\vbox to #1{\hsize#2\vfill}\vrule}\hrule}\hfill}%
    \else
    {\hfill\epsfbox{#3}\hfill}%
    \fi}

\begin{document}

\pagestyle{plain}

\begin{flushleft}
\babar-\BaBarType-\BaBarYear/\BaBarNumber \\
SLAC-PUB-\SLACPubNumber\\
arXiv:1810.04724 [hep-ex]\\
\end{flushleft}

\title{{\large \bf Search for a Stable Six-Quark State at \babar}}

%
\author{J.~P.~Lees}
\author{V.~Poireau}
\author{V.~Tisserand}
\affiliation{Laboratoire d'Annecy-le-Vieux de Physique des Particules (LAPP), Universit\'e de Savoie, CNRS/IN2P3,  F-74941 Annecy-Le-Vieux, France}
\author{E.~Grauges}
\affiliation{Universitat de Barcelona, Facultat de Fisica, Departament ECM, E-08028 Barcelona, Spain }
\author{A.~Palano}
\affiliation{INFN Sezione di Bari and Dipartimento di Fisica, Universit\`a di Bari, I-70126 Bari, Italy }
\author{G.~Eigen}
\affiliation{University of Bergen, Institute of Physics, N-5007 Bergen, Norway }
\author{D.~N.~Brown}
\author{Yu.~G.~Kolomensky}
\affiliation{Lawrence Berkeley National Laboratory and University of California, Berkeley, California 94720, USA }
\author{M.~Fritsch}
\author{H.~Koch}
\author{T.~Schroeder}
\affiliation{Ruhr Universit\"at Bochum, Institut f\"ur Experimentalphysik 1, D-44780 Bochum, Germany }
% ...
\collaboration{The \babar\ Collaboration}
\noaffiliation

\begin{abstract}
Recent investigations have suggested that the six-quark combination $uuddss$ could be a deeply bound state ($S$) that has 
eluded detection so far, and a potential dark matter candidate. We report the first search for a stable, doubly strange 
six-quark state in $\Upsilon \rightarrow S \bar\Lambda \bar\Lambda$ decays based on a sample of $90 \times 10^6\, \Upsilon(2S)$ 
and $110 \times 10^6\, \Upsilon(3S)$ decays collected by the \babar\ experiment.  No signal is observed, and 90\% 
confidence level limits on the combined $\Upsilon(2S,3S) \rightarrow S \bar\Lambda \bar\Lambda$ branching fraction 
in the range $(1.2-1.4)\times 10^{-7}$ are derived for $m_S < 2.05 \GeV$. These bounds set stringent limits on the 
existence of such exotic particles.
\end{abstract}


\maketitle

\setcounter{footnote}{0}

A new stable state of matter may still be undiscovered. While the vast majority of known hadrons can be 
described as either quark-antiquark or three-quark combinations, other multi-quark possibilities are allowed 
by quantum chromodynamics (QCD). Among those, the six-quark configuration $uuddss$ is of particular 
interest, as its spatial wave function is completely symmetric, and generic arguments imply that it should 
be the most tightly bound six-quark state (see, e.g., Ref.~\cite{Preskill:1980mz}). This property was already noticed by 
Jaffe 40 years ago~\cite{Jaffe:1976yi}. He predicted the existence of a loosely bound $uuddss$ state with a mass close 
to $2150 \MeV$~\cite{units}, dubbed the H-dibaryon. As its mass is above the $m_p +m_e + m_\Lambda = 2055\MeV$ threshold, 
the H-dibaryon would have a typical weak interaction lifetime. Numerous negative experimental results were taken as 
evidence against such a particle, including observations of doubly strange hypernuclei decays~\cite{Ahn:2001sx,Takahashi:2001nm}, 
searches for narrow $\Lambda p \pi^-$ resonances in $\Upsilon$ decays~\cite{Kim:2013vym} and direct searches 
for new neutral particles (see e.g. Ref.~\cite{Badier:1986xz, Bernstein:1988ui, Belz:1995nq,AlaviHarati:1999ds,Gustafson:1976hd}).

The situation is markedly different if the potential is deeply attractive, as advocated by Farrar~\cite{Farrar:2017eqq}. Below 
$2055 \MeV$, the $uuddss$ configuration acquires a cosmological lifetime, as its decay would have to proceed via doubly-weak 
interactions, and it is absolutely stable if it is lighter than $2(m_p + m_e) = 1878 \MeV$. Such a bound state, tentatively named 
$S$~\cite{Naming}, hasn't been excluded so far by hypernuclei decays or direct searches for long-lived neutral states (the latter were 
limited to masses above $\sim2 \GeV$ due to the large neutron background~\cite{Gustafson:1976hd}). While current lattice QCD calculations 
suggest a small value of the $uuddss$ binding energy, at the level of ${\cal O}(10) \MeV$~\cite{Beane:2012vq,Francis:2018qch}, lattice 
systematic uncertainties remain too large to rule out a deeply bound state.

Although not all authors agree (see e.g. Ref.~\cite{Gross:2018ivp,Kolb:2018bxv,McDermott:2018ofd}), a stable six-quark state might 
also have cosmological implications. If dark matter is composed of nearly equal numbers of $u,d$, and $s$ quarks, its formation rate 
is driven by the quark-gluon plasma transition to the hadronic phase and the quark and anti-quark abundances. As the same 
source is also responsible for determining the residual amount of ordinary matter in the universe, this framework would explain 
both the dark matter density and the baryon asymmetry, two seemingly unrelated quantities. A specific realization of this 
scenario, six-quark dark matter with $m_S \sim 1860-1880 \MeV$, can reproduce the observed ratio of dark matter to ordinary 
matter densities within $\sim15\%$~\cite{Farrar:2018hac}. 

Being a flavor singlet, the $S$ particle does not couple to pions or other mesons. The $S$-nucleon interaction 
cross-section is expected to be suppressed compared to that of nucleon-nucleon interactions, and its production 
rate is several orders of magnitude below that for neutrons. Given that a low-mass $S$ is 
difficult to kinematically distinguish from a neutron, these attributes might explain why this state has 
escaped detection so far. Despite these difficulties, several search strategies have been proposed. 
Among them, the exclusive decay $\Upsilon \rightarrow S \bar\Lambda \bar\Lambda$~\cite{cc} stands out for 
its simplicity and robustness. The short-distance nature of the gluonic source increases the overlap with 
the compact $S$ wave function, enhancing its production rate compared to other mechanisms involving 
baryons. Heuristic arguments suggest an inclusive six-quark production rate in $\Upsilon(1S,2S,3S)$ 
decays at the level of $10^{-7}$, albeit with significant uncertainties~\cite{Farrar:2017eqq}. No specific prediction 
for the exclusive $S \bar\Lambda \bar\Lambda$ final state has been made so far, though this channel could 
conceivably account for a large fraction of the total production rate. 

We report herein the first search for a stable, doubly strange six-quark configuration produced in $\Upsilon(2S,3S)$ 
decays~\cite{stable}. For completeness, we probe the entire mass range compatible with a stable state: 
$0 \gev < m_S < 2.05 \gev$. The analysis is based on a sample containing $90 \times 10^6 \, \Upsilon(2S)$ and 
$110 \times 10^6 \, \Upsilon(3S)$ decays collected with the \babar\ detector at the \pep2\ asymmetric-energy 
$\epem$ collider operated at the SLAC National Accelerator Laboratory. The integrated luminosities of the $\Upsilon(2S)$ 
and $\Upsilon(3S)$ samples are 14\invfb and 28\invfb, respectively~\cite{Lees:2013rw}. Additional samples of 428\invfb 
collected at the $\Upsilon(4S)$ peak, as well as in the vicinity of the $\Upsilon(2S,3S)$ resonances, are used 
to estimate the background. The \babar\ detector is described in detail elsewhere~\cite{Bib:Babar,TheBABAR:2013jta}. 
To avoid experimental bias, we examine the data signal region only after finalizing the analysis strategy.

Simulated events are used to optimize the selection procedure and assess the signal efficiency. Signal events 
are generated for $0 \gev < m_S < 2.2\GeV$ in steps of $0.2 \GeV$. The $S$ angular distribution 
is simulated using an effective Lagrangian based on a constant matrix element for the different 
arrangements of angular momentum between the final state particles, assuming that angular momentum suppression 
effects are small~\cite{PrivateFarrar} (see appendix for a detailed description). A second model 
based on a phase space distribution is used to assess systematic uncertainties. The interaction between six-quark states 
and matter is expected to be similar to that of neutrons, albeit with reduced cross-sections. For the purpose of simulating 
the signal, we model these interactions similarly to those of neutrons. As an extreme alternative, we simulate six-quark states 
as non-interacting particles, and we assign the difference between these two models as a systematic uncertainty. To study 
the background, we generate generic $\Upsilon(2S,3S,4S)$ decays with EvtGen~\cite{Lange:2001uf}, while the continuum 
$\epem \rightarrow \q\overline{q}$ ($q=u,d,s,c$) background is estimated using a data-driven approach described below. The 
detector acceptance and reconstruction efficiencies are determined using a Monte Carlo (MC) simulation based on 
GEANT4~\cite{Bib::Geant}. Time-dependent detector inefficiencies and background conditions, as monitored during data-taking 
periods, are included in the simulation. 

\begin{figure}[htb]
\begin{center}
  \includegraphics[width=0.49\textwidth]{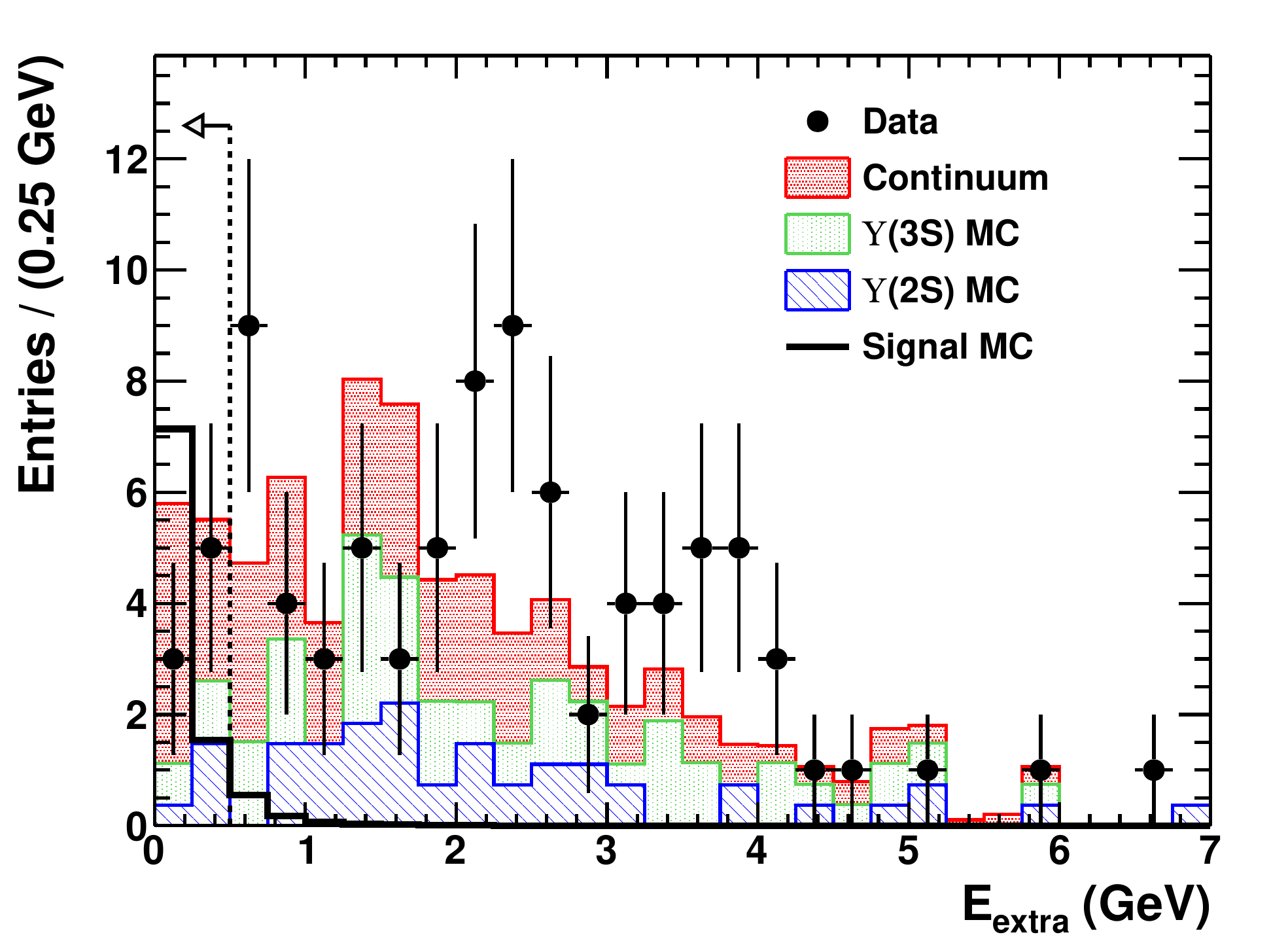} 
\end{center}
\caption
{The distribution of the extra neutral energy, $E_{\rm extra}$, before performing the kinematic fit for the combined 
$\Upsilon(2S)$ and $\Upsilon(3S)$ data sets, together with various background estimates (stacked histograms) and signal 
MC predictions (solid line). The requirement on $E_{\rm extra}$ is indicated by a dashed line. The signal MC is normalized to a 
branching fraction $\BR(\Upsilon \rightarrow S \bar\Lambda \bar\Lambda) = 5\times10^{-7}$. The background MC normalization 
is described in the text.}
\label{Fig1}
\end{figure}

\begin{figure}[htb]
\begin{center}
  \includegraphics[width=0.49\textwidth]{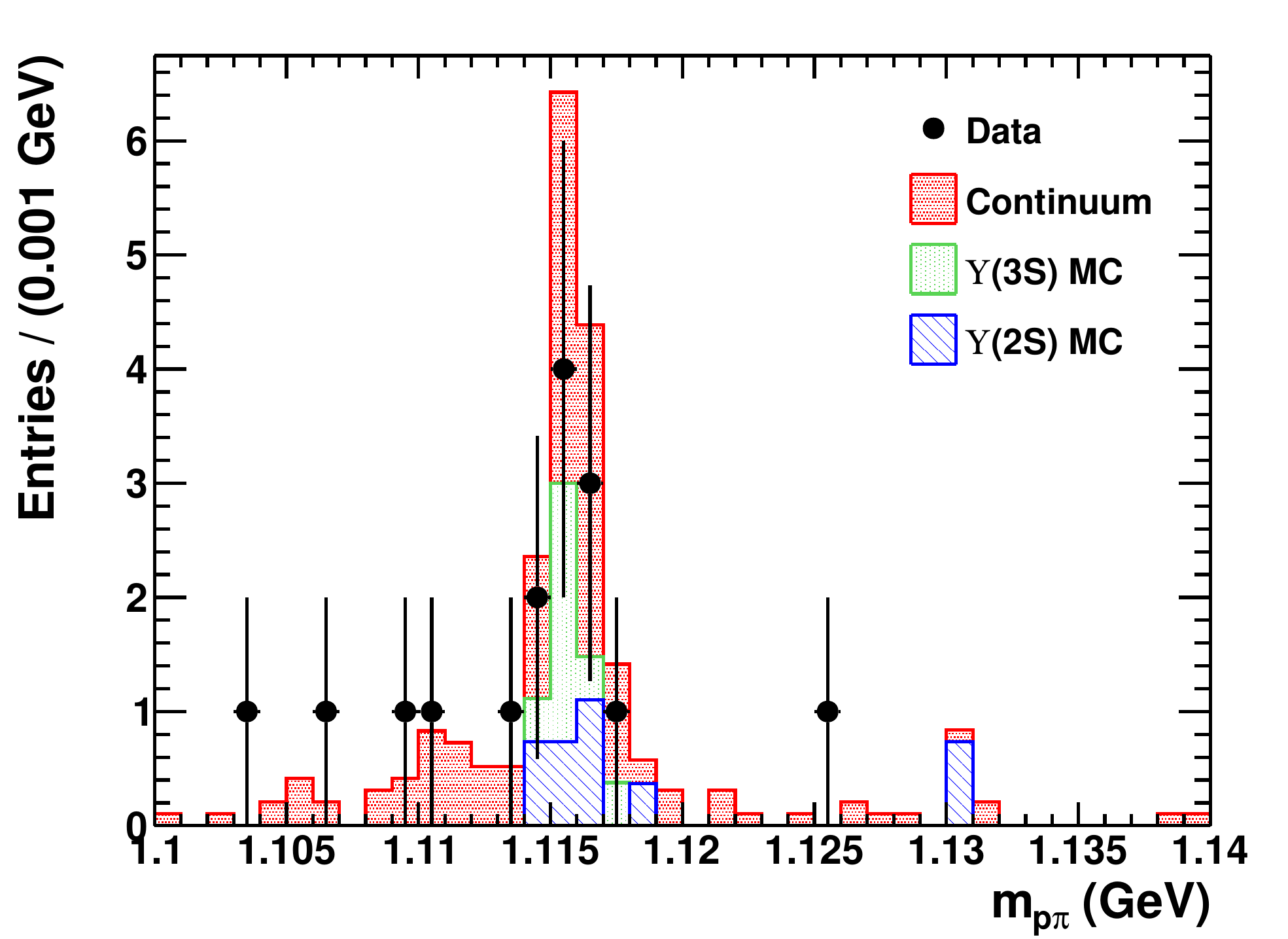} 
\end{center}
\caption
{The distribution of the $p\pi$ invariant mass, $m(p\pi)$, before performing the kinematic fit for the combined 
$\Upsilon(2S)$ and $\Upsilon(3S)$ data sets, together with various background estimates (stacked histograms). Two 
entries per event are plotted.}
\label{Fig2}
\end{figure}

We select events containing at most five tracks and two $\Lambda$ candidates with the same strangeness, reconstructed in the 
$\Lambda \Lambda\rightarrow p\pi^- p \pi^-$ final state with $1.10 \GeV < m_{p \pi} < 1.14 \GeV$. 
One additional track not associated with a $\Lambda$ candidate with a distance of closest approach from the primary interaction 
point (DOCA) larger than 5 cm is allowed to account for particles produced from secondary interactions with the detector 
material. The (anti)protons must be selected by particle identification (PID) algorithms. This requirement, which is 
approximately 95\% efficient for identifying both protons and antiprotons, removes a large amount of background from 
four-pion final states. To further improve the signal purity, the $\Lambda$ flight vector is measured as the distance 
between the primary interaction point and the $\Lambda$ decay vertex. The flight significance of each $\Lambda$ candidate, 
defined as the length of this vector dived by its uncertainty, must be larger than 5. The cosine of the angle between 
the $\Lambda$ momentum and the flight vector must also be greater than 0.9. In addition, the total energy of clusters 
in the electromagnetic calorimeter not associated with charged particles, $E_{\rm extra}$, 
must be less than $0.5 \gev$. To account for possible interactions between the $S$ candidate and 
the calorimeter, the sum excludes clusters that are closer than an angle of 0.5 rad to the inferred $S$ direction. 
Moreover, the distance between the cluster and the proton is required to be greater than 40 cm to reduce the 
contribution of cluster fragments. The $E_{\rm extra}$ distribution after applying all other selection criteria is 
shown in Fig.~\ref{Fig1}. The background is dominated by hadronic events containing several strange baryons and additional 
charged and neutral particles, a fraction of which escapes undetected. The selection procedure is tuned to maximize the 
signal sensitivity, taking into account the systematic uncertainties related to $S$ production and interaction with 
detector material in the calculation. The $p \pi^-$ mass distribution obtained after applying these criteria is shown 
in Fig.~\ref{Fig2}. A total of 8 $\Upsilon \rightarrow S \bar\Lambda \bar\Lambda$ candidates are selected.

\begin{figure}[htb]
\begin{center}
  \includegraphics[width=0.49\textwidth]{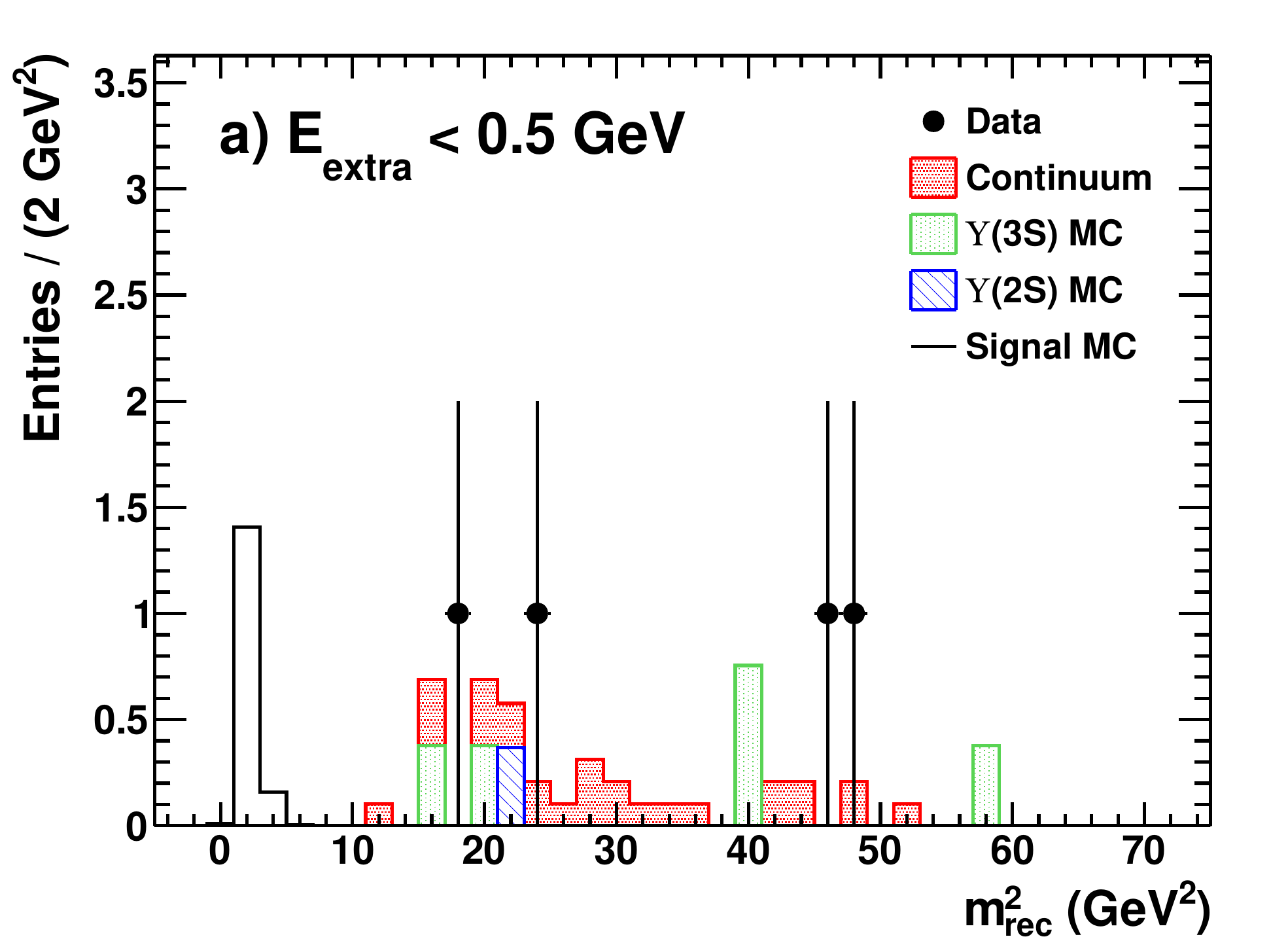} \\
  \includegraphics[width=0.49\textwidth]{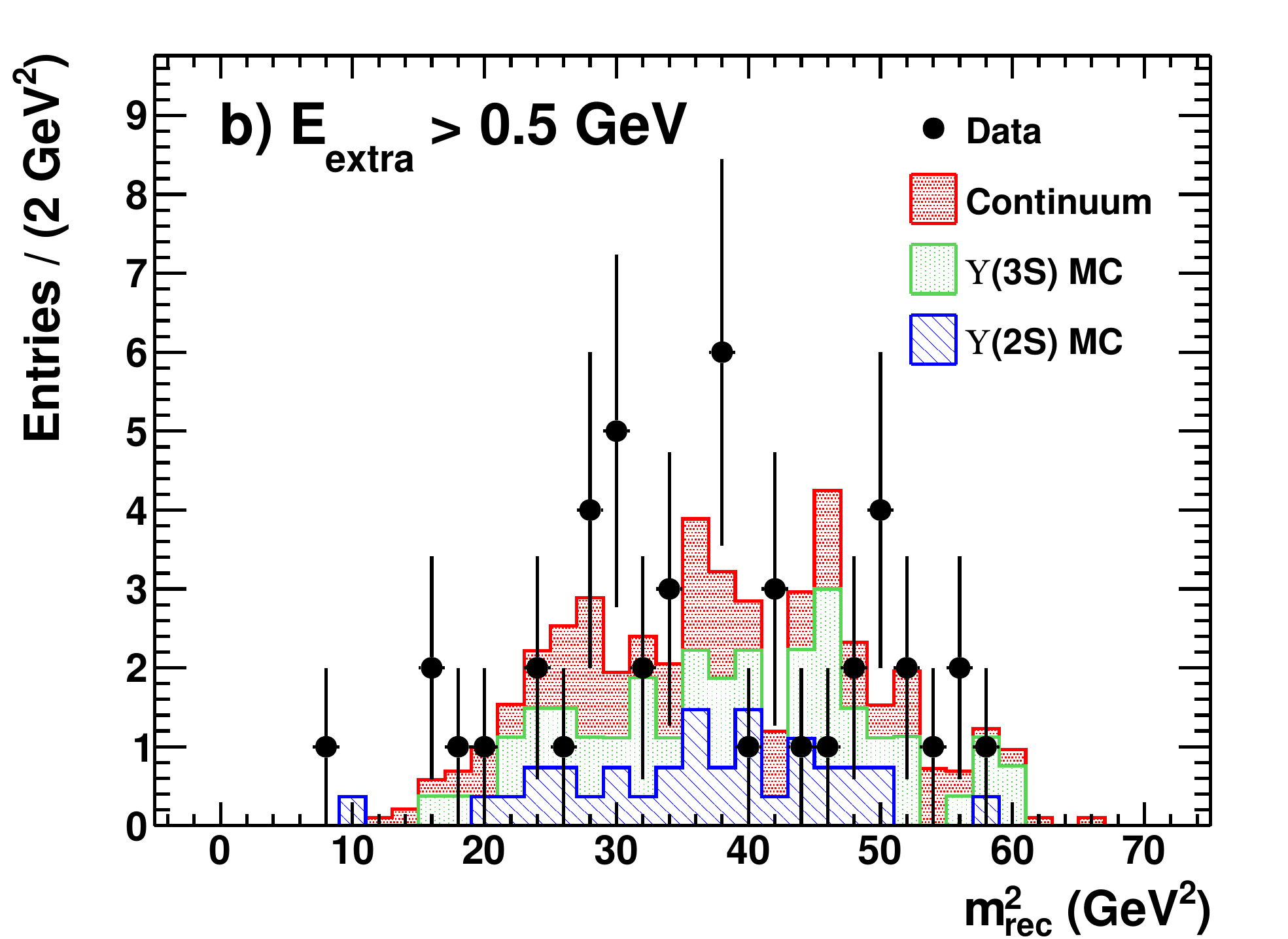} 
\end{center}
\caption
{The distribution of the recoil mass squared against the $\Lambda \Lambda$ system, $m_{\rm rec}^2$, after 
performing the kinematic fit for the combined $\Upsilon(2S)$ and $\Upsilon(3S)$ data sets, together with 
various background estimates (stacked histograms) and a signal example for (a) the $E_{\rm extra}< 0.5 \GeV$ 
signal region and (b) the $E_{\rm extra} >  0.5 \GeV$ sideband data sample. }
\label{Fig3}
\end{figure}

The events are then fit, imposing a mass constraint to each $\Lambda$ candidate and requiring a common origin, compatible 
with the beam interaction point within its uncertainty. We select combinations with $\chi^2<25$ (for 8 d.o.f.), retaining 
half of the previously selected candidates. The signal is identified as a peak in the recoil mass squared against the 
$\Lambda \Lambda$ system, $m_{\rm rec}^2$, in the region $ 0 \GeV^2 \lesssim  m_{\rm rec}^2 \lesssim 5 \GeV^2$. The recoil 
mass squared allows for negative values arising from the limited resolution on the reconstructed $\Lambda$ candidates, 
providing a better estimator of the efficiency near $m_S \sim 0 \GeV$ than the recoil mass. The $m_{\rm rec}^2$ distribution 
is shown in Fig.~\ref{Fig3}a, together with various background predictions and a simulated signal assuming 
$m_S = 1.6 \gev$ and a branching fraction $\BR(\Upsilon \rightarrow S \bar\Lambda \bar\Lambda) = 1\times10^{-7}$. No events 
are observed in the signal region, and the expected background is found to be negligible as well.

The continuum $\epem \rightarrow \q\overline{q}$ ($q=u,d,s,c$) background is estimated from the data collected at the 
$\Upsilon(4S)$ peak. This data sample contains contributions from both continuum and $\Upsilon(4S)$ events. The 
latter is evaluated from the generic $\Upsilon(4S)$ MC sample and found to be negligible, as those decays tend to have 
higher multiplicity and are much more suppressed than continuum production by our selection. The data collected at the 
$\Upsilon(4S)$ resonance are therefore a good representation of the continuum background. 

The $\Upsilon(2S,3S)$ background components are estimated from the corresponding MC simulations. The contributions are normalized using 
sideband data obtained by applying all the selection criteria previously described but requiring $E_{\rm extra}$ to be 
greater than $0.5 \GeV$ instead of below that threshold. The $\Upsilon(2S,3S)$ MC components are found to underestimate 
the observed $p \pi$ yield, and we adjust their overall normalizations to improve the agreement with the data. The 
resulting correction factors, determined to be respectively 1.25 and 1.55 for the $\Upsilon(2S)$ and $\Upsilon(3S)$ samples, are 
propagated throughout the analysis.   

A data-driven estimate of the background is also derived from the sideband data. Similarly to signal events, this sample contains 
predominantly two real $\Lambda$ particles with additional (undetected) particles. Since the difference in $E_{\rm extra}$ is essentially 
due to the interaction of those particles with the calorimeter, sideband data provide a good approximation of the expected background 
in the signal region. The corresponding recoil mass distribution is displayed in Fig.~\ref{Fig3}b. Similarly to the background estimate 
previously described, sideband data predict a negligible level of background in the signal region.

The efficiency as a function of the $S$ mass is derived from the corresponding MC sample. For each mass hypothesis, 
we define a signal region in the $m_{\rm rec}^2$ distribution as the symmetric interval around the nominal $S$ mass 
containing 99\% of the reconstructed $S$ candidates. Its typical size is of the order of $2.5 \GeV^2$. The efficiency 
varies between 7.2\% near threshold to 8.2\% near $m_S = 2\GeV$. It is mainly driven by the detector acceptance and 
the $\Lambda \rightarrow p \pi$ branching fraction.

The main uncertainties on the efficiency arise from the modeling of the $\Upsilon \rightarrow S \bar\Lambda \bar\Lambda$ 
angular distribution and the limited knowledge of the $S$-matter interactions. The former varies between 4\% to 15\%, 
assessed by taking the difference between the predictions based on the simplified Lagrangian to those obtained using a phase 
space distribution for $\Upsilon$ decays. The latter is estimated by considering the difference between simulations modeling 
the $S$ as a neutron or a non-interacting 
particle. The corresponding uncertainty ranges from 8\% to 10\%. A systematic uncertainty of 8\% is included 
to account for the difference in $\Lambda$ reconstruction efficiencies between data and MC calculations, determined from control samples 
in data~\cite{Aubert:2007uf}. Both the uncertainty on the $\Lambda \rightarrow p \pi$ branching fraction 
(1.6\%~\cite{PDG}) and the finite MC sample size ($\sim 1.5\%$) are also propagated. 

\begin{figure}
\begin{center}
\includegraphics[width=0.49\textwidth]{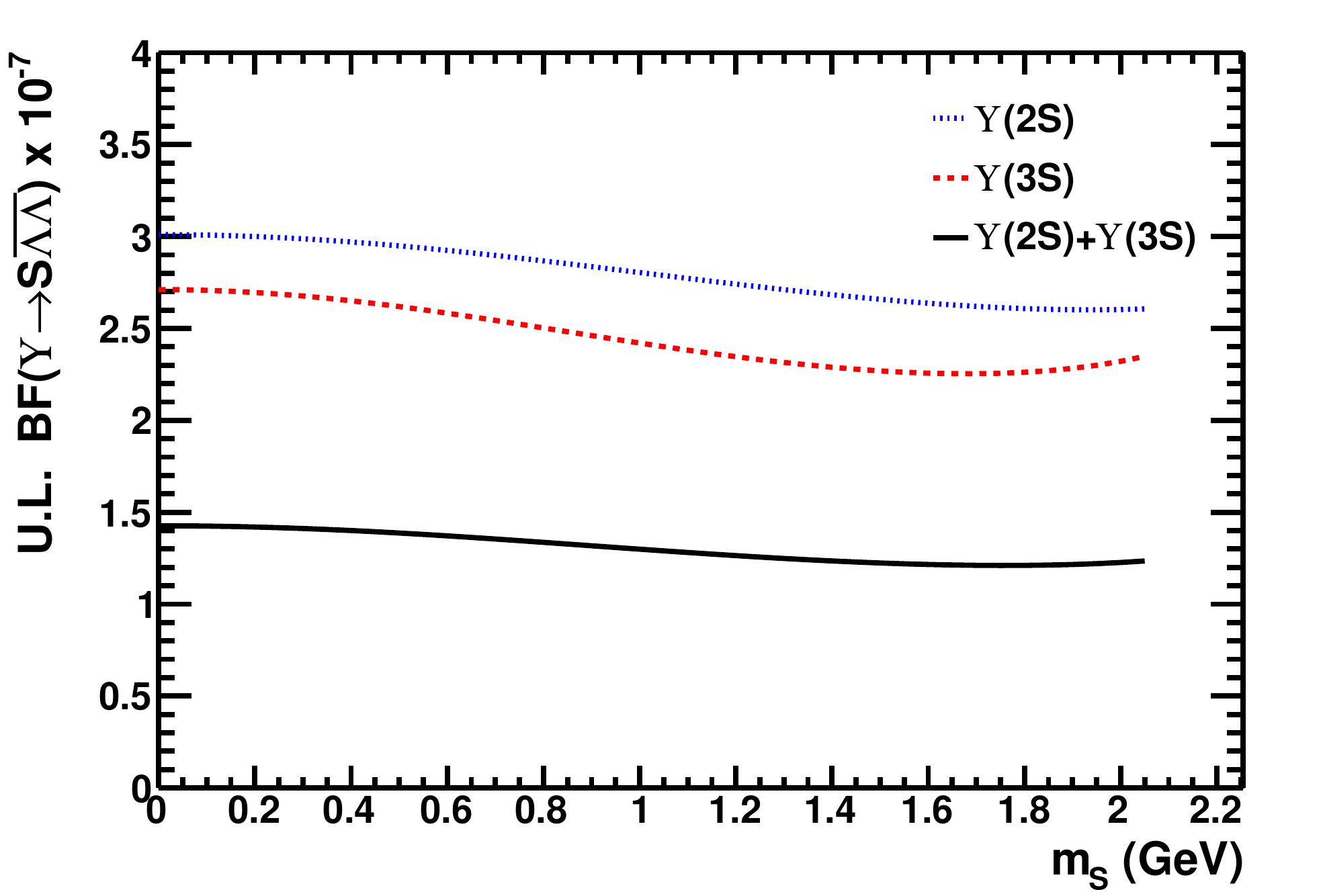} 
\end{center}
\caption
{The 90\% CL upper limits on the $\Upsilon(2S,3S) \rightarrow S \bar\Lambda \bar\Lambda$ branching fraction for the $\Upsilon(2S)$ 
and $\Upsilon(3S)$ data sets, as well as the combined sample assuming the same partial width.}
\label{Fig4}
\end{figure}

No significant signal is observed, and we derive 90\% confidence level (CL) upper limits on the $\Upsilon(2S,3S) \rightarrow S \bar\Lambda \bar\Lambda$ 
branching fractions, scanning $S$ masses in the range $0 \GeV < m_S < 2.05\GeV$ in steps of $50 \MeV$ (approximately half the 
signal resolution). For each mass hypothesis, we evaluate the upper bound on the number of signal events from the $m_{\rm rec}^2$ 
distribution with a profile likelihood method~\cite{Rolke:2004mj}. This approach treats the background as a Poisson process whose 
unknown mean is estimated from the number of observed background events, set to zero in this instance. Systematic uncertainties are 
included by modeling the signal efficiency as a Gaussian distribution with the appropriate variance. In addition to the contributions 
previously described, the limits include an additional uncertainty of 0.6\% associated with the uncertainty on the number of 
$\Upsilon(2S)$ and $\Upsilon(3S)$ decays. The results are shown in Fig.~\ref{Fig4} for the $\Upsilon(2S)$ and $\Upsilon(3S)$ data 
sets, as well as the combined sample assuming the same partial width.

In conclusion, we performed the first search for a stable $uuddss$ configuration in $\Upsilon$ decays. No signal is observed, and 
90\% CL limits on the combined $\Upsilon(2S,3S) \rightarrow S \bar\Lambda \bar\Lambda$ branching fraction of $(1.2-1.4)\times 10^{-7}$ 
are derived for $ m_S < 2.05\GeV$. These results set stringent bounds on the existence of a stable, doubly 
strange six-quark state.

\section{Acknowledgments}
We thank G. Farrar for providing us with the theoretical model to simulate the signal and useful discussions. 
We are grateful for the 
extraordinary contributions of our \pep2\ colleagues in
achieving the excellent luminosity and machine conditions
that have made this work possible.
The success of this project also relies critically on the 
expertise and dedication of the computing organizations that 
support \babar.
The collaborating institutions wish to thank 
SLAC for its support and the kind hospitality extended to them. 
This work is supported by the
US Department of Energy
and National Science Foundation, the
Natural Sciences and Engineering Research Council (Canada),
the Commissariat \`a l'Energie Atomique and
Institut National de Physique Nucl\'eaire et de Physique des Particules
(France), the
Bundesministerium f\"ur Bildung und Forschung and
Deutsche Forschungsgemeinschaft
(Germany), the
Istituto Nazionale di Fisica Nucleare (Italy),
the Foundation for Fundamental Research on Matter (The Netherlands),
the Research Council of Norway, the
Ministry of Education and Science of the Russian Federation, 
Ministerio de Econom\'{\i}a y Competitividad (Spain), the
Science and Technology Facilities Council (United Kingdom),
and the Binational Science Foundation (U.S.-Israel).
Individuals have received support from 
the Marie-Curie IEF program (European Union) and the A. P. Sloan Foundation (USA). 


\section{Appendix}
The $S$ angular distribution is simulated using the following amplitude:
\begin{widetext}
\begin{eqnarray*}
|A|^2 & = & 2\frac{mM(m^2 - \alpha) - M^2\beta + 2(mM + \alpha - \beta)(\beta -\gamma -mM)  - mM(\gamma-m^2) + m^2(\alpha - \beta + \gamma - M_Y^2)}{(m^2 - M^2 - 2 \alpha + M_Y^2)(m^2-M^2 - 2\gamma + M_Y^2)} \\
      &   & +\frac{2m^2(M^2+m^2-2\alpha + M_Y^2) - 2mM(m^2 - \alpha-\beta + \gamma) + 2(m^2 - \alpha)(\beta - \gamma)- \beta(M_Y^2+m^2-M^2 - 2\alpha)}{(m^2 - M^2 - 2 \alpha + M_Y^2)^2}\\
      &   & +\frac{2m^2(M^2-m^2-2\gamma+M_Y^2) + 2mM(\alpha -\beta+\gamma-m^2)      - 2(m^2-\gamma)(\alpha - \beta)     - \beta(M_Y^2-m^2-M^2 - 2\gamma)  }{(m^2 - M^2 - 2 \gamma + M_Y^2)^2}
\end{eqnarray*}
\end{widetext}
where $\alpha = p \cdot q$, $\beta = p \cdot p'$, $\gamma = p' \cdot q$, $q$ is the 4-momentum of the 
$\Upsilon(2S,3S)$, $p$ ($p'$) is the 4-momentum of the first (second) $\Lambda$, $m$ is the $\Lambda$ 
mass and $M$ is an effective mass, taken to be $m_\Lambda$.


\begin{thebibliography}{9}
\bibitem{units} Natural units ($\hbar=c=1$) are used throughout this paper.
\bibitem{cc} Charge conjugation is implied throughout the text.
\bibitem{stable} The search is also applicable to any scenario with a long-lived six-quark state escaping the detector.
\end{thebibliography}
\end{document}